\documentclass[conference]{IEEEtran}
\usepackage{graphicx}
\usepackage{caption}
\usepackage{subcaption}

\usepackage{longtable}
\usepackage{stfloats}
\usepackage{graphicx}

\usepackage{booktabs}
\usepackage{multirow}

\usepackage{ifpdf}
\usepackage[T1]{fontenc}
\usepackage{amssymb}
\usepackage{cite}

\ifCLASSINFOpdf
   \usepackage{graphicx}
	\usepackage{epstopdf}
   \graphicspath{{images/}}   
   \graphicspath{{../pdf/}{../jpeg/}{../eps/}}
   \DeclareGraphicsExtensions{.pdf,.jpeg,.png,.eps}
\else
   \usepackage[dvipdfm]{graphicx}
   \usepackage{bmpsize}
   \graphicspath{{../eps/}}
   \DeclareGraphicsExtensions{.eps}
\fi
\graphicspath{{images/}}

\usepackage[cmex10]{amsmath}
\usepackage{multirow}

\usepackage{multirow}

\usepackage{booktabs}

\usepackage{paralist}

\usepackage{multicol}

\usepackage[shortcuts,acronym, nomain]{glossaries}

\makeglossaries 

\newacronym{rnn}{RNN}{Recurrent Neural Network}
\newacronym{awgn}{AWGN}{Additive White Gaussian Noise}
\newacronym{gru}{GRU}{Gated Recurrent Unit}
\newacronym{lstm}{LSTM}{Long-Short Term Memory}
\newacronym{snr}{SNR}{Signal to Noise Ratio}
\newacronym{urllc}{URLLC}{Ultra Reliable Low Latency Communication}
\newacronym{dl}{DL}{Deep Learning}
\newacronym{ml}{ML}{Machine Learning}
\newacronym{qos}{QoS}{Quality of Service}
\newacronym{mmtc}{MMTC}{Massive Machine Type Communications}
\newacronym{iot}{IoT}{Internet of Things}
\newacronym{cnn}{CNN}{Convolutional Neural Network}
\newacronym{gan}{GAN}{Generative Adversarial Network}
\newacronym{dbn}{DBN}{Deep Belief Network}
\newacronym{phy}{PHY}{Physical Layer}
\newacronym{ann}{ANN}{Artificial Neural Network}
\newacronym{mlp}{MLP}{Multi Layer Perceptron}
\newacronym{ldpc}{LDPC}{Low-Density Parity-Check}
\newacronym{ber}{BER}{Bit Error Rate}
\newacronym{gpu}{GPU}{Graphical Processing Unit}
\newacronym{rsc}{RSC}{Recursive Systematic Convolutional}
\newacronym{map}{MAP}{Maximum A Posteriori}
\newacronym{adam}{ADAM}{Adaptive Moment Estimation}
\newacronym{qpsk}{QPSK}{Quadrature Phase Shift Keying}
\newacronym{qam}{QAM}{Quadrature Amplitude Modulation}
\newacronym{agv}{AGV}{Automatic Guided Vehicle}
\newacronym{atp}{ATP}{Automatic Train Pairing}
\newacronym{osi}{OSI}{Open Systems Interconnect}
\newacronym{uwb}{UWB}{Ultra-Wide Band}
\newacronym{ai}{AI}{Artificial Intelligence}
\newacronym{e2e}{E2E}{End-to-End}

\begin{document}

\title{AI-assisted PHY technologies for 6G and beyond wireless networks}

%
\author{\IEEEauthorblockN{Raja Sattiraju, Andreas Weinand and Hans D. Schotten}
\IEEEauthorblockA{Chair for Wireless Communication \& Navigation \\
University of Kaiserslautern\\
\{sattiraju, weinand, schotten\}@eit.uni-kl.de}}
\maketitle

\begin{abstract}
\ac{ml} and \ac{ai} have become alternative approaches in wireless networks beside conventional approaches such as model based solution concepts. Whereas traditional design concepts include the modelling of the behaviour of the underlying processes, \ac{ai} based approaches allow to design network functions by learning from input data which is supposed to get mapped to specific outputs (training). Additionally, new input/output relations can be learnt during the deployement phase of the function (online learning) and make \ac{ai} based solutions flexible, in order to react to new situations. Especially, new introduced use cases such as \ac{urllc} and \ac{mmtc} in 5G make this approach necessary, as the network complexity is further enhanced compared to networks mainly designed for human driven traffic (4G, 5G xMBB). The focus of this paper is to illustrate exemplary applications of \ac{ai} techniques at the \ac{phy} of future wireless systems and therfore they can be seen as candidate technologies for e.g. 6G systems.

\end{abstract}

\section{Introduction}


In order deliver an optimal user experience, future networks need to pro-actively learn the interests and requirements of network stakeholders and adapt their behaviour autonomously at any given time. The main challenge here is to assist the network in order to orchestrate itself with an optimal configuration in the sense of \ac{e2e} performance metrics. This can be achieved based on  learning new situations and their consequences on these metrics. \ac{ml}, as one of the most powerful artificial intelligence tools, constitutes a promising solution in order to tackle this challenge. \ac{ml} techniques have been applied to various domains such as computer vision, natural language processing, social network filtering, drug design, and many more, where they have produced results comparable to and in some cases superior to human experts.\

In the field of wireless networking, \ac{ml} techniques were initially applied to upper layers (e.g. resource management\cite{Challita2017}). However, recently the focus has been shifted to using \ac{ml} at \ac{phy} layer since this basically eliminates the need for any traditional signal processing to be applied apriori. Some of the examples include obstacle detection \cite{Sattiraju2017a} and localization\cite{Vieira2017}, channel coding\cite{Ortuno1992,Sattiraju2018}, modulation recognition\cite{OShea2017}, physical layer security \cite{Weinand2017_2}, and channel estimation and equalization\cite{Wen2015}. The motivation for the \ac{ml} approach is two fold - 1.\ac{ml} based approaches, are designed to achieve an optimal \ac{e2e} performance, whereas traditional signal processing is done by logically separated blocks that are independently optimized, and 2.\acp{ann} are shown to be universal function approximators\cite{Hornik1989} and are known to be Turing complete. Hence, these algorithms can be executed faster and at lower energy cost (ca n be parallelized on distributed memory architectures)



In this paper, we examplary analyse the possibility of using \ac{ml} at \ac{phy} layer by means of three different candidate applications, Channel coding using \acp{rnn}, Ranging \& Obstacle detection using supervised learning algorithms and \ac{phy} layer security using unsupervised learning algorithms.

\section{\ac{ml} at \ac{phy} - Applications}
Due to the enormous variety of \ac{phy} layer technologies existing, we have choosen three exemplary applications for \ac{ml} solutions at the \ac{phy} layer. These are introduced in the following and supported by initial evaluations of the performance of the respective \ac{ml} approaches.

\subsection{Channel Coding using \acp{rnn}}
\begin{figure}{b!}
\centering
\includegraphics[width=\linewidth]{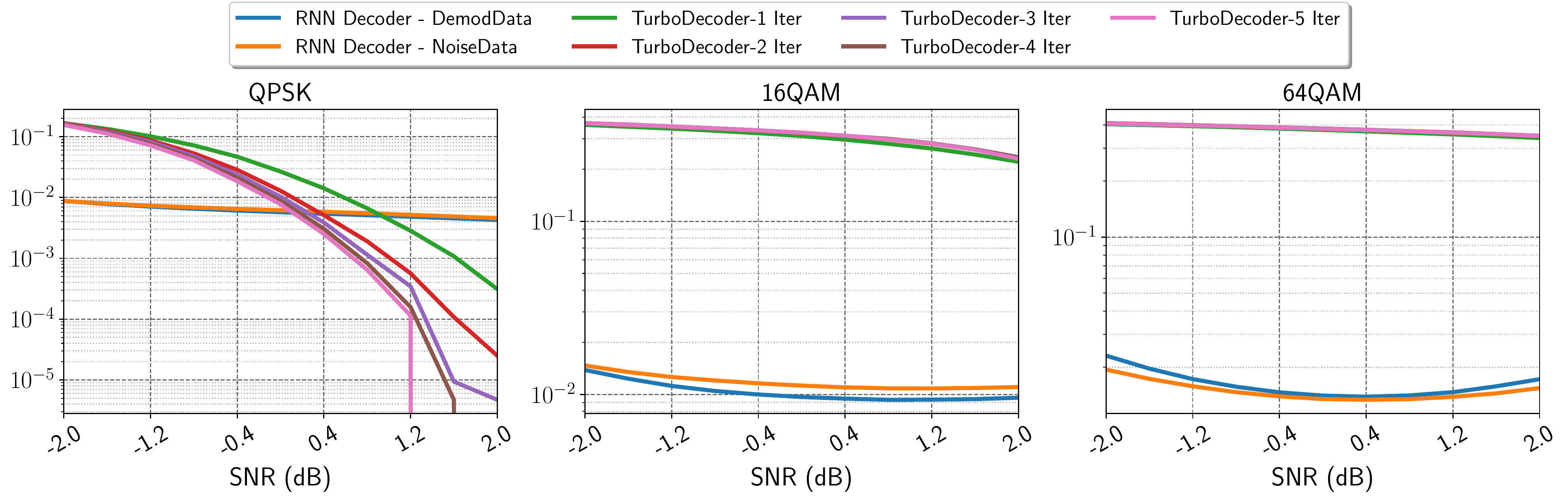}
\caption{RNN decoder vs. turbo decoder for different modulation orders}
\label{coding}
\end{figure}

Channel Coding has been one of the central disciplines driving the success stories of current generation LTE systems and beyond. In particular, turbo codes are mostly used for cellular and other applications where a reliable data transfer is required for latency-constrained communication in the presence of data-corrupting noise. However, the decoding algorithm for turbo codes is computationally intensive, thereby limiting its applicability in hand-held devices. In \cite{Sattiraju2018}, we study the feasibility of using  \ac{dl} architectures based on \acp{rnn} for encoding and decoding of turbo codes. Simulation results (Fig. \ref{coding}) show, that the proposed \ac{rnn} model outperforms the decoding performance of a conventional turbo decoder at low \ac{snr} regions.

\subsection{Ranging \& Obstacle Detection using Supervised Learning}
\begin{figure}{}
	\centering
	\includegraphics[width=0.85\linewidth]{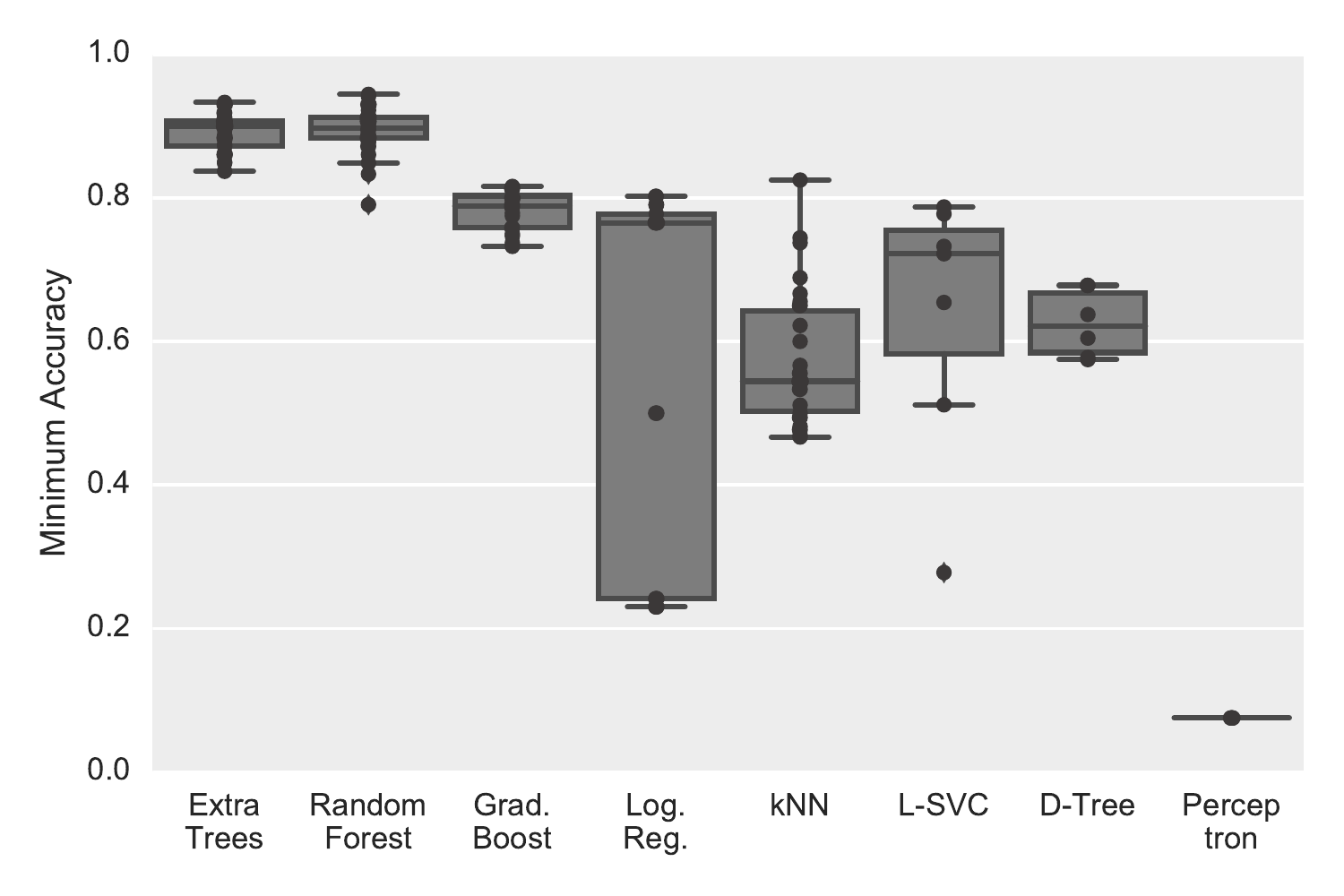}
	\caption{Performance of various supervised learning algorithms}
	\label{gridsearch}
\end{figure}

Short Range wireless devices are becoming more and more popular for ubiquitous sensor and actuator connectivity in industrial communication scenarios. Apart from communication only scenarios, there are also mission-critical use cases where the distance between the two communicating nodes needs to be determined precisely. Applications such as \acp{agv}, \ac{atp} additionally requires the devices to scan the environment and detect any potential humans/obstacles. \ac{uwb} has emerged as a promising candidate for real-time ranging and localization due to advantages such as large channel capacity, better co-existence with legacy systems due to low transmit power, better performance in multipath environments etc. In \cite{Sattiraju2017a}, the raw time domain \ac{uwb} waveforms is used in order to detect obstacles by constructing a multiclass hypothesis and using supervised learning for predictions. Simulation results show (Fig. \ref{gridsearch}) that the Ensemble tree based methods are able to calculate the likelihood of obstacle collision with accuracies close to 95\%.

\subsection{\ac{phy} Layer Security using Unsupervised Learning}
\begin{figure}{}
\begin{subfigure}{.25\textwidth}
  \centering
  \includegraphics[width=\linewidth]{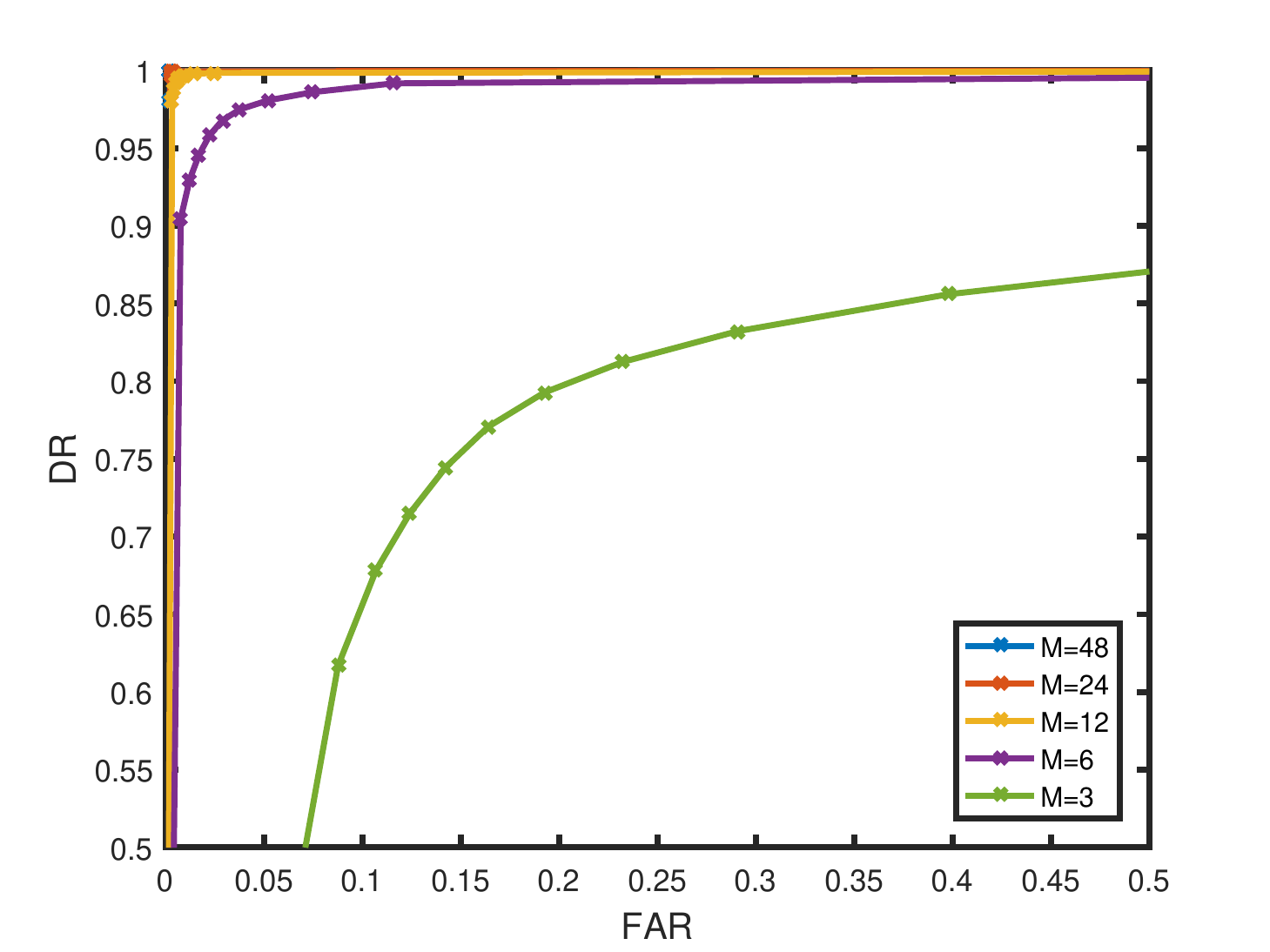}
  \caption{Linear scale}
  \label{fig:sfig1}
\end{subfigure}%
\begin{subfigure}{.25\textwidth}
  \centering
  \includegraphics[width=\linewidth]{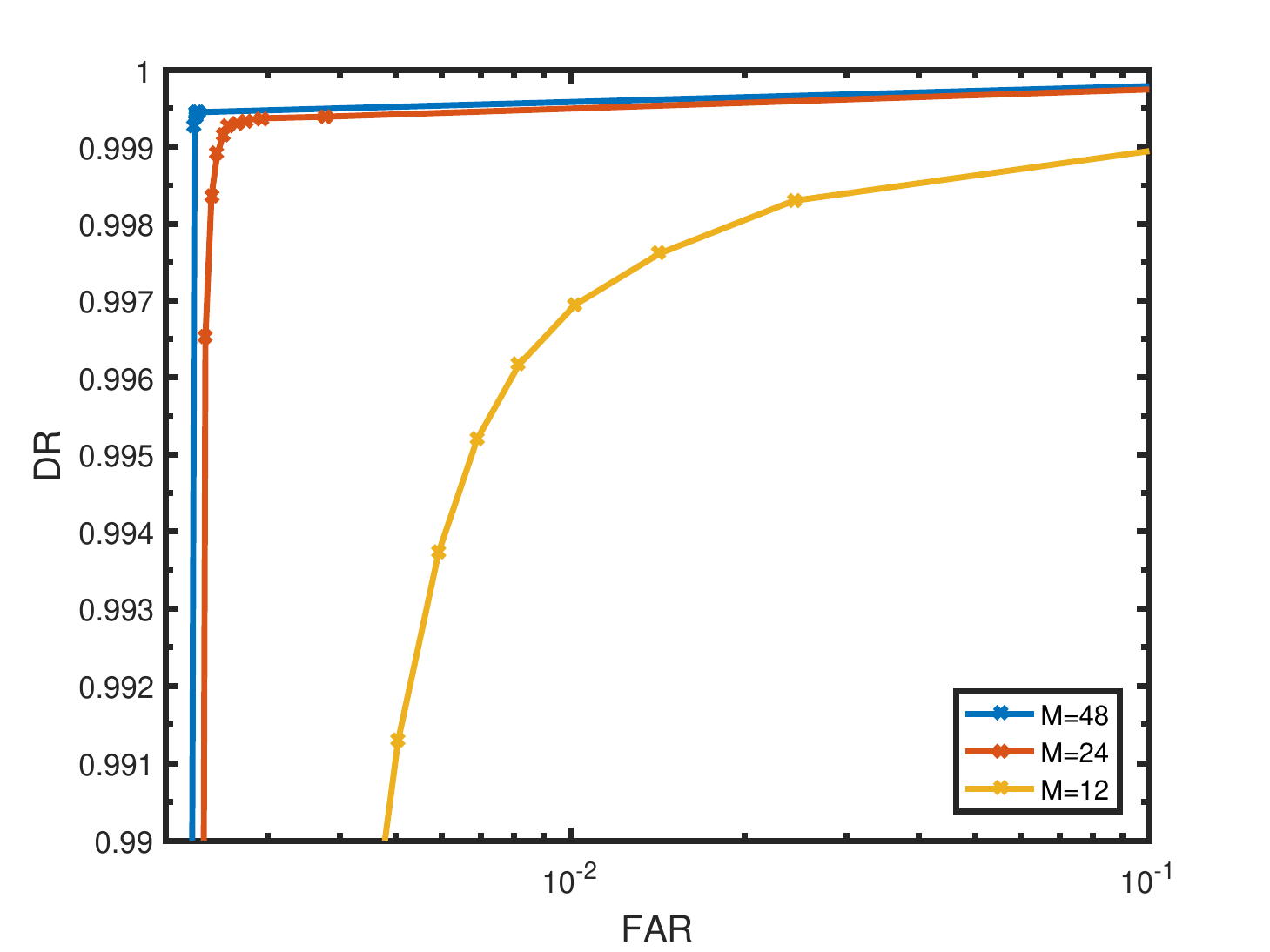}
  \caption{Log scale}
  \label{fig:sfig2}
\end{subfigure}
\caption{GMM based method with different feature dimensions}
\label{physec}
\end{figure}

An alternative approach, compared to conventional schemes such as certificates or authentication tags, in order to guarantee message authenticity and integrity is to take non-cryptographic information, such as protocol metadata, into account. Due to spoofability of metadata on the logical level, such as frame counters or traffic patterns, physical layer protocol metadata, e.g. received signal strength or channel estimation data, is more suited for that purpose. Especially the channel impulse or frequency response has been used as a feature in numerous works in order to indicate the origin of transmitted data packets, e.g. in \cite{Baracca.2012}.
If a receiver needs to authenticate a packet from a specific transmitter, he checks whether the respective channel estimation matches with the previous ones based on received packets of that user. This can be achieved by different methods, either by conventional statistics such as generalized likelihood ratio testing \cite{Baracca.2012} or \ac{ml} based approaches such as SVMs \cite{Pei.2014} or GMM \cite{Weinand2017_1}.  
Fig. \ref{physec} shows the performance of the GMM based approach in \cite{Weinand2017_1} (detection rate DR vs. false alarm rate FAR). Depending on the number of subcarriers $M$ (pilot tones) that are used for channel estimation in an OFDM system, the performance increases or decreases.

\section{Conclusion}
The results from our investigated applications show us, that these might be good candidates for AI-assisted 6G technologies. Further investigations are necessary, e.g. evaluating the scenarios using more different \ac{ml} techniques and combining data driven AI decisions with semantic information in order to achieve more reliable decisions (reinforcement learning). 

\bibliographystyle{IEEEtran}
\bibliography{6g_summit}

\end{document}